\documentclass[prd,aps,nofootinbib]{revtex4}
\usepackage{epsfig}
\usepackage{amsfonts}
\usepackage{rotating}
\usepackage{sparticles}
\usepackage{dcolumn}
\usepackage{bm}
\usepackage{hyperref}
\hypersetup{
    bookmarks=true,         
    unicode=false,          
    pdftoolbar=true,        
    pdfmenubar=true,        
    pdffitwindow=false,     
    pdfstartview={FitH},    
    pdftitle={My title},    
    pdfauthor={Author},     
    pdfsubject={Subject},   
    pdfcreator={Creator},   
    pdfproducer={Producer}, 
    pdfkeywords={keyword1} {key2} {key3}, 
    pdfnewwindow=true,      
    colorlinks=true,       
    linkcolor=red,          
    citecolor=cyan,        
    filecolor=magenta,      
    urlcolor=green,           
    linktocpage=true
}
\usepackage{amsmath,amssymb,amsthm,graphicx,latexsym}
\usepackage{subfigure}
\usepackage{pstricks}
\usepackage{color}

\usepackage{mathrsfs}

\newcommand{\be}{\begin{equation}}
\newcommand{\ee}{\end{equation}}
\newcommand{\bea}{\begin{eqnarray}}
\newcommand{\eea}{\end{eqnarray}}
\newcommand{\bml}{\begin{subequations}}
\newcommand{\eml}{\end{subequations}}
\newcommand{\bfig}{\begin{figure}}
\newcommand{\efig}{\end{figure}}

\newcommand{\slashed}{\hspace{-1.1ex}/}

\begin{document}

\title{A step towards exploring the features of Gravidilaton sector in Randall-Sundrum scenario via lightest Kaluza-Klein graviton mass}
\author{{\textcolor{black}{Sayantan Choudhury}}$^{1}$\footnote{Electronic address: {sayanphysicsisi@gmail.com}} ${}^{}$ and {\textcolor{black}{Soumitra SenGupta}}$^{2}$
\footnote{Electronic address: {tpssg@iacs.res.in}} ${}^{}$}
\affiliation{$^1$Physics and Applied Mathematics Unit, Indian Statistical Institute, 203 B.T. Road, Kolkata 700 108, India}
\affiliation{$^2$Department of Theoretical Physics,
Indian Association for the Cultivation of Science,
2A \& 2B Raja S.C. Mullick Road,
Kolkata - 700 032, India.}

\begin{abstract}
In this paper we study the role of the 5D Gauss-Bonnet corrections and two loop higher genus contribution to the gravity
action in type IIB string theory inspired low energy supergravity theory in the light of
 gravidilatonic interactions on the lightest Kaluza-Klein graviton mass spectrum.
 From the latest constraints on the lightest Kaluza-Klein graviton mass as obtained from the 
ATLAS dilepton search in 7 TeV proton-proton collision, we have shown that due to the presence of Gauss-Bonnet
 and string loop corrections, the warping solution in an ${\bf AdS_{5}}$ bulk is quite distinct from
Randall-Sundrum scenario. We discuss the constraints on the model parameters to fit with present ATLAS data.
\end{abstract}


\maketitle

Search for extra dimensions in Large Hadron Collider (LHC) experiments via Kaluza-Klein (KK) graviton mode is an extensive area of collider
research. In particular the recent ATLAS experiment put some stringent lower bound on the lightest KK graviton mass in the context of
Randall-Sundrum warped geometry model via the dilepton decay of the KK graviton. Randall-Sundrum model becomes phenomenologically popular because of its
promise to resolve the fine tuning problem in connection with Higgs mass without introducing any hierarchical parameter. This model is defined on a slice of ${\bf AdS_{5}}$
with the bulk being an Einstein-anti-desitter spacetime. Recent conflict between the ATLAS data and Randall-Sundrum model in estimating the lightest KK graviton mass
motivate us here to extend the bulk beyond 
Einstein-anti-desitter to a string loop corrected Gauss-Bonnet anti-desitter space and explore the graviton search experiment again to look for a possible stringy signature in collider
physics. 

In this work we have first explored the phenomenological features of string modified warped geometry
 in presence of 5D Gauss-Bonnet coupling and gravidilaton coupling in a 5D bulk.
 Here the 5D warped geometry model
has been proposed by making use of the following sets of assumptions as a building block:
\begin{itemize}
 \item The Einstein's gravity sector is modified by the introduction of Gauss-Bonnet correction 
\cite{Choudhury:2012yh,Choudhury:2012kw,Choudhury:2013qza,Choudhury:2013eoa,Choudhury:2013dia,Choudhury:2013yg,Choudhury:2014hna,Kim:1999dq,Lee:2000vf,Kim:2000pz,Kim:2000ym} and string two loop 
correction \cite{Choudhury:2013eoa,Choudhury:2013dia,Choudhury:2013yg} originated from holographic dual ${\bf CFT_{4}}$
 disk amplitude in {\it type II B string theory} or its {\it low energy supergravity theory} \cite{Green1,Green2,Polchinski1,Polchinski2}.
\item  The well known ${\bf S^{1}/Z_{2}}$ orbifold compactification is considered.
\item We considered that the system is embedded in 5D AdS bulk where the background warped metric has a Randall-Sundrum (RS) like structure
 with ${\bf AdS_{5}\times S^{5}}$ geometry \cite{Randall:1999ee,Randall:1999vf}.
\item  The compactification radius/modulus
is assumed to be independent of four dimensional coordinates (by Poincare invariance) as well as extra dimensional coordinate \cite{Choudhury:2013yg}.
\item The strength of the gravidialton interaction is determined by dilaton degrees of freedom which are assumed to be confined within the bulk.
\item Additionally, the dilaton field also interact with the 5D bulk cosmological constant $\Lambda_{5}$ via dilaton coupling.
\item The Higgs field is localized at the visible brane and the hierarchy problem is resolved via Planck to TeV scale warping.
\item The modulus can be stabilized by introducing
scalar in the ${\bf AdS_{5}}$ bulk without any fine tuning following Goldberger-Wise mechanism \cite{Goldberger:1999uk,Goldberger:1999un,Goldberger:1999wh,Choudhury:2014hna}.
 \item It is assumed that the requirement of the solution
     of the gauge hierarchy problem (or equivalently naturalness problem/fine
     -tuning problem) is still obeyed as this resolution was one of the main goal
     of involving such warped geometry model in the perturbative limit of our proposed setup.
     
\item Additionally while determining the value of the model parameters from the proposed setup we also require
     that the bulk curvature is less than the five dimensional Planck scale $M_{5}$ so that
     the classical solution can be trusted \cite{Davoudiasl:1999jd,Das:2013lqa,Choudhury:2014sua}.
\end{itemize}
In the present article first we compute the warping solution in presence of 5D Gauss-Bonnet as well as gravidilaton coupling and the
two loop higher genus string loop correction. Further using the solution we have discussed the detailed phenomenological
features of lightest Kaluza-Klein graviton mass in the light of the constraint obtained from the ATLAS dilepton search.
 We further compare our results with
 the results obtained from the well known Randall-Sundrum model and comment on the present status of both of them in the light of 
present collider constraints. In this analysis we use the combined phenomenological 
bounds on Gauss-Bonnet coupling $\alpha_{5}$ obtained from Higgs diphoton and dilepton decay channels \cite{ATLAS:2013mma} and from Higgs mass
 from the ATLAS \cite{Aad:2012tfa} and CMS \cite{Chatrchyan:2013lba} data within $5\sigma$ C.L.. This bound lies below the upper bound of viscosity-entropy ratio \cite{Brigante:2007nu} and satisfies
the unitarity bound \cite{Brigante:2007nu,Cremonini:2011iq,Brigante:2008gz,Buchel:2009tt,Ge:2008ni,Ge:2009eh,Hofman:2009ug} on the GB coupling.
We have also discussed the explicit dependence and the phenomenological feature of the lightest Kaluza-Klein graviton mass on the 5D Gauss-Bonnet coupling, gravidilaton coupling and the
two loop higher genus string loop correction by scanning our analysis throughout the allowed parameter space in the perturbative regime of the proposed setup.

We start our discussion with the following 5D action of the two brane warped geometry model given by \cite{Choudhury:2013yg}:
\begin{widetext}
\be\begin{array}{llll}\label{eq1}
 \displaystyle S=\int d^{5}x \left[\sqrt{-g_{(5)}}\left\{\frac{M^{3}_{5}}{2}R_{(5)}+\frac{\alpha_{5}M_{5}}{2}(1-A_{1}e^{\theta_{1}\phi(y)})
\left[R^{ABCD(5)}R^{(5)}_{ABCD}-4R^{AB(5)}R^{(5)}_{AB}+R^{2}_{(5)}\right]\right.\right.\\ \left.\left.
\displaystyle ~~~~~~~~~~~~~~~~~~~~~~~~~~~~~~~~~~~~~~~~~~~~~~~~~~~~~~+\frac{g^{AB(5)}}{2}\partial_{A}\phi(y)\partial_{B}\phi(y)-2\Lambda_{5}e^{\theta_{2}\phi(y)}\right\}+
\displaystyle \sum^{2}_{i=1}\sqrt{-g^{(i)}_{(5)}}\left[{\cal L}^{field}_{i}-V_{i}\right]\delta(y-y_{i})\right]
\end{array}\ee
\end{widetext}
with $A,B,C,D=0,1,2,3,4(extra~dimension)$ and a conformal two-loop string coupling constant $A_{1}$.
 Here $i$ signifies the brane index, $i=1(\text{hidden})$, $2(\text{visible})$ and
${\cal L}^{field}_{i}$ is the Lagrangian for the fields on the ith brane where ith the brane tension $V_{i}$ and 
 $\phi(y)$ represent the dilaton field which is dynamical in the bulk with respect to the extra dimensional coordinate `y'. 
The background metric describing slice of the ${\bf AdS_{5}}$ is given by,
\be\begin{array}{llllll}\label{brane}
   \displaystyle ds^{2}_{5}=g_{AB}dx^{A}dx^{B}=e^{-2A(y)}\eta_{\alpha\beta}dx^{\alpha}dx^{\beta}+r^{2}_{c}dy^{2}
   \end{array}\ee
where $r_{c}$ is the dimensionless quantity in the Planckian unit representing the compactification radius of extra dimension.
Here the orbifold points are $y_{i}=[0,\pi]$
and periodic boundary condition is imposed in the closed interval $-\pi\leq y\leq\pi$.

Varying the action stated in equation(\ref{eq1}) and neglecting the back reaction of all the other brane/bulk fields except gravity and dilaton,
the five dimensional Bulk Einstein's equation turns out to be,
\begin{widetext}
\be\begin{array}{lllll}\label{eneqn}
    \displaystyle \sqrt{-g_{(5)}}\left[G^{(5)}_{AB}+\frac{\alpha_{5}}{M^{2}_{5}}\left(1-A_{1}e^{\theta_{1}\phi(y)}\right)H^{(5)}_{AB}\right]
=-\frac{e^{\theta_{2}\phi(y)}}{M^{3}_{5}}\left[\Lambda_{5} \sqrt{-g_{(5)}}g^{(5)}_{AB}+\sum^{2}_{i=1}V_{i}\sqrt{-g^{(i)}_{(5)}}
g^{(i)}_{\alpha\beta}\delta^{\alpha}_{A}\delta^{\beta}_{B}\delta(y-y_{i})\right]
   \end{array}\ee
\end{widetext}
where the five dimensional Einstein's tensor and the Gauss-Bonnet tensor is given by:
\begin{widetext}
\be\begin{array}{llll}\label{et}
    G^{(5)}_{AB}=\left[R^{(5)}_{AB}-\frac{1}{2}g^{(5)}_{AB}R_{(5)}\right],
   \end{array}\ee

\be\begin{array}{llll}\label{gbp}
  H^{(5)}_{AB}=2R^{(5)}_{ACDE}R_{B}^{CDE(5)}-4R_{ACBD}^{(5)}R^{CD(5)}
-4R_{AC}^{(5)}R_{B}^{C(5)}+2R^{(5)}R_{AB}^{(5)}\\ ~~~~~~~~~~~~~~~~~~~~~~~~~~~~~~~~~~~~~~~~~~~-\frac{1}{2}g^{(5)}_{AB}
\left(R^{ABCD(5)}R^{(5)}_{ABCD}-4R^{AB(5)}R^{(5)}_{AB}+R^{2}_{(5)}\right).
   \end{array}\ee
\end{widetext}

Similarly varying equation(\ref{eq1}) with respect to the dilaton field the gravidilaton equation of motion turns out to be:
\begin{widetext}
\be\begin{array}{llll}\label{jk11}
\displaystyle \frac{\theta_{2}}{M^{2}_{5}}\sum^{2}_{i=1}V_{i}\sqrt{-g^{(i)}_{(5)}}e^{\theta_{2}\phi(y)}\delta(y-y_{i})
=\sqrt{-g_{(5)}}\left\{\alpha_{5}A_{1}\theta_{1}\left[R^{ABCD(5)}R^{(5)}_{ABCD}-4R^{AB(5)}R^{(5)}_{AB}+R^{2}_{(5)}\right]
\right.\\ \left.~~~~~~~~~~~~~~~~~~~~~~~~~~~~~~~~~~~~~~~~~~~~~~~~~~~~~~~~~~~~~~~~~~~~~~~~~~~~~~~~~~~~~~~~~~~~~~~~~~~~~~~~~~\displaystyle 
+2\frac{\Lambda_{(5)}}{M^{2}_{5}}\theta_{2}e^{\theta_{2}\phi(y)}+\frac{\Box_{(5)} \phi(y)}{M_{5}}\right\}
   \end{array}\ee
\end{widetext}
where the five dimensional D'Alembertian operator is defined as:
 \be\Box_{(5)}\phi(y)=\frac{1}{\sqrt{-g_{(5)}}}\partial_{A}\left(\sqrt{-g_{(5)}}\partial^{A}\phi(y)\right).\ee
To solve equation(\ref{eneqn}) and equation(\ref{jk11}) we assume that the dilaton is 
weakly coupled to gravity (weak coupling $\theta_{1}$) and the bulk cosmological constant (weak coupling $\theta_{2}$)  since 
the Gauss-Bonnet coupling is an outcome of perturbative correction to gravity at the quadratic order.
Now including the well known ${\bf Z_{2}}$ orbifolding symmetry at the orbifold points, $y_{i}=[0,\pi]$, 
for perturbative regime of solution due to the presence of very weak couplings $\theta_{1}$, $\theta_{2}$ and $\alpha_{(5)}$  
we can neglect the contribution from first two terms in the right hand side of Eq~(\ref{jk11}) in the bulk. The contribution from the left hand side in Eq~(\ref{jk11})
autometically vanishes within bulk. Finally we are left with only the last term in the right hand side of Eq~(\ref{jk11}) from which  
we get the following solution of the dilaton degrees of freedom within the bulk \cite{Choudhury:2013yg}:
\be\begin{array}{llll}\label{gradilatonicAS}
 \displaystyle  \phi(y)=c_{1}|y|+c_{2}
   \end{array}\ee
where $c_{1}$ and $c_{2}$ are the integration constants to be determined from the value of $\phi(y)$ at the boundaries.
We write the dimensionless exponent of the 
dilaton factors by substituting Eq~(\ref{gradilatonicAS}) at the orbifold point $y_{i}=\pi$:
\be\begin{array}{llll}\label{exp1}
    \displaystyle \chi_{1}=\theta_{1}\phi(\pi)=\theta_{1}(c_{1}|\pi|+c_{2}),\\
\displaystyle \chi_{2}=\theta_{2}\phi(\pi)=\theta_{2}(c_{1}|\pi|+c_{2}).
   \end{array}\ee
where we redefine the exponents by using the symbols $\chi_{1}$ and $\chi_{2}$.
In the present context we have chosen that the two different dilaton couplings are connected through, $\theta_{1}=-\theta_{2}$ for which we have:
\be\begin{array}{llll}\label{exp2}
    \displaystyle \chi_{1}=-\chi_{2}=\theta_{1}(c_{1}|\pi|+c_{2}).
   \end{array}\ee
For numerical estimations we take the dilation couplings $\theta_{1},\theta_{2}$ to be small to keep it within the perturbative regime of solution
and we set the arbitrary integration
constants $c_{1},c_{2}$ to a desired value for which the dimensionless exponents of the dilaton factors are fixed at:
\be \lim_{\theta_{1}\rightarrow weak}e^{-\chi_{1}}=\lim_{\theta_{1}\rightarrow weak}
e^{\chi_{2}}=e^{11}.\ee 

 
Such a value of the dimensionless exponent of the 
dilaton factor produce a large enhancement even for small value of dilaton coupling parameter $\theta_{1},\theta_{2}$ and moderate values for and $\phi(0)$ and $\phi(\pi)$. 
In our subsequent calculation this 
enhancement factor will play a significant role. As we will see later such a choice is inspired from the requirement of Planck to TeV scale warping as well as to keep
the mass of the first excited state of the Kaluzu-Klein mode graviton above the bound set by LHC which is $1.01$~TeV as can be seen from the Table~(\ref{tab1}). Thus
this choice sets a bound on the dilaton coupling consistent with LHC constraint. 

In presence of dilaton the solution of the five dimensional bulk {\it Einstein Gauss Bonnet} equation of motion
 at leading order in GB coupling ($\alpha_{5}$) turns out to be \cite{Choudhury:2013yg}:
\begin{widetext}
\be\begin{array}{lll}\label{eq2}
    \displaystyle A(y)=k_{\pm}r_{c}|y|\\ \displaystyle~~~~~~\displaystyle =\sqrt{\frac{3M^{2}_{5}}{16\alpha_{5}(1-A_{1}e^{\theta_{1}\phi(y)})}
\left[1\pm\sqrt{\left(1+\frac{4\alpha_{5}(1-A_{1}e^{\theta_{1}\phi(y)}) \Lambda_{5}e^{\theta_{2}\phi(y)}}{9M^{5}_{5}}\right)}\right]}r_{c}|y|.
   \end{array}\ee
\end{widetext}
\begin{table*}
\centering
\begin{tabular}{|c|c|c|c|c|c|c|c|}
\hline
\hline
\hline
 & &  & Amount of & & Lower limit   & Lower limit & Lower limit  \\
 & &  & warping to & & of the  & of the  & of the\\
 & &5D   & accommodate & & lightest KK   & lightest KK & lightest KK\\
& &Planck & {\bf SM Higgs} & & graviton mass  &  graviton mass &  graviton mass\\
& & mass& {\bf on the}  & & from the & from the & from the\\
{\large$\epsilon_{RS}=\frac{k_{RS}}{M_{5}}$} & {\large $\epsilon_{\bf M}=\frac{k_{\bf M}}{M_{5}}$} & & {\bf visible brane} & $Z_{\bf T}$
&{\bf LHC }  & {\bf proposed } & {\bf Randall- }\\
&  &  &  & & {\bf ATLAS }  & {\bf theoretical }& {\bf Sundrum (RS) }\\
& & $M_{5}$ &{\large $e^{-k_{\bf M}r_{c}\pi}$} & & {\bf dilepton } & {\bf model} & {\bf model}\\
& & (using Eq~(\ref{fivedx})) & (using Eq~(\ref{massasdphi})) & &{\bf search} & & \\
& &  &  & &{(within $5\sigma$ C.L.)}   & $m^{G}_{1}$ & $m^{G,{\bf RS}}_{1}$\\
& &  &  & &$m^{G,{\bf ATLAS}}_{1}$  & (using Eq~(\ref{jkq3z}))& ($\times 10^{-1}$)\\
& & (in~ $M_{Pl}$) & ($\times 10^{-17}$) & & (in~ TeV) & (in~ TeV)& (in~ TeV)\\
\hline\hline\hline
0.01 & 2.45 & 1.56 &  0.801 & 0.015 & 1.01& 1.68 & 0.22
\\
0.03 & 7.34 & 2.71 & 0.461  & 0.081 & 1.48& 5.04 & 0.46
\\
0.05 &12.23 & 3.50 & 0.357 & 0.175 & 1.88& 8.41 & 0.65
\\
 0.07 & 17.13 & 4.14 & 0.302  & 0.290 & 2.04 & 11.77 & 0.81
\\
 0.09 & 22.02 & 4.69 & 0.267 & 0.422 & 2.17 & 15.13 & 0.96 \\
 0.10 & 24.47 & 4.95 & 0.253 & 0.495 & 2.22 & 16.82 & 1.03
\\
\hline
\hline
\hline
\end{tabular}
\vspace{.4cm}
\caption{\label{tab1}
 Comparitive study between the lower limit of the lightest Kaluza-Klein graviton mass for $n=1$ mode from the proposed theoretical model, the well known
Randall-Sundrum (RS) model and the LHC ATLAS dilepton serach in 7 TeV
proton-proton collision
. Here to study the outcome from our proposed setup we fix the model parameters 
as, Gauss-Bonnet coupling $\alpha_{5}=5\times 10^{-7}$ (which is consistent with the solar system constraint \cite{Chakraborty:2012sd}, combined constraint from the
 Higgs mass and favoured decay channels $H\rightarrow (\gamma\gamma,\tau\bar{\tau})$ \cite{ATLAS:2013mma} using ATLAS \cite{Aad:2012tfa} and CMS \cite{Chatrchyan:2013lba} data) and string two-loop correction 
$A_{1}=0.05$ with $M_{Pl}\approx 10^{19}GeV$ and Higgs mass $m_{H}=125$ GeV (within the $5\sigma$ statistical C.L. of ATLAS and CMS). Throughout the analysis additionally we have maintained another constraint between the gravidilaton coupling and the dilaton coupling with
 the 5D cosmological constant in AdS space-time, $\theta_{2}=-\theta_{1}$. This implies at the leading order approximation, 
$\theta_{2}\phi(\pi)=-\theta_{1}\phi(\pi)=11$ at the visible brane.} 
\end{table*}

\begin{figure*}[htb]
\centering
\includegraphics[width=12.5cm,height=10cm]{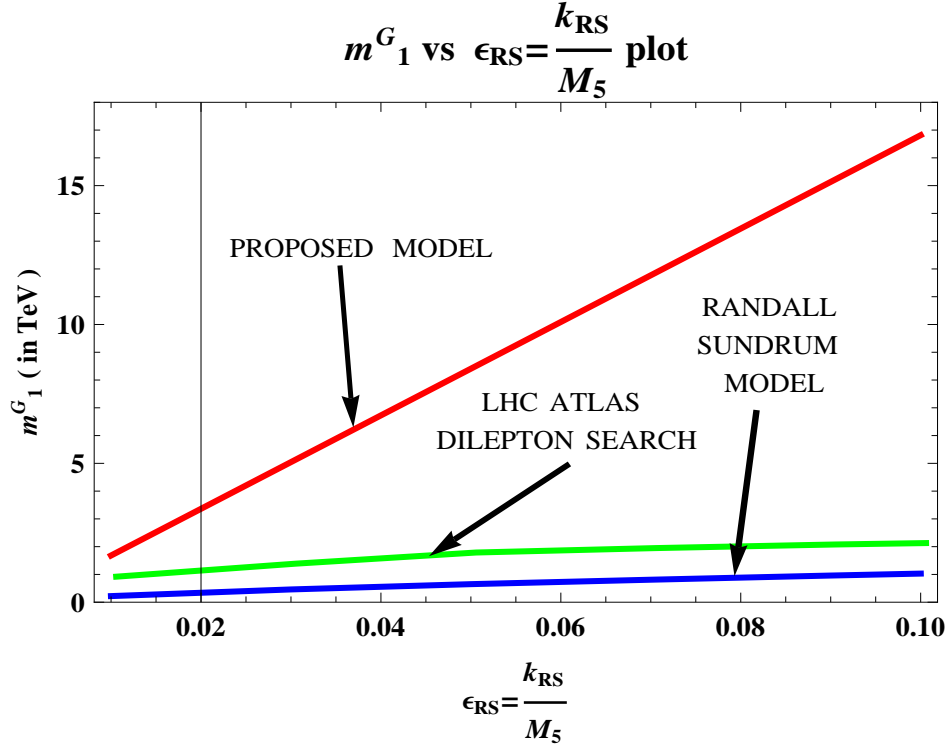}
\caption{Variation of the lightest Kaluza-Klein graviton
 mass $m^{G}_{1}$ for $n=1$ mode from the proposed model (red curve) and Randall Sundrum (RS) model (blue curve) with respect to 
the phenomenological parameter {\large$ \epsilon_{RS}=\frac{k_{RS}}{M_{5}}$} within the range $0.01<\epsilon_{RS}<0.10$. We have also shown the present status
of the lower limit of the lightest Kaluza-Klein graviton
 mass for LHC ATLAS dilepton search by the green curve as depicted in table~(\ref{tab1}).
 Here for this plot we fix the model parameters as, $\alpha_{5}=5\times 10^{-7}$, $A_{1}=0.05$,
$\theta_{2}=-\theta_{1}$, $\theta_{2}\phi(\pi)\sim 11$ and $\theta_{1}\phi(\pi)\sim -11$. Here the allowed region is in the upper half  
of the green curve. The rest of the region (below the geen curve) is ruled out. }
\label{f1}
\end{figure*}

\begin{figure*}[htb]
\centering
\includegraphics[width=12.5cm,height=8cm]{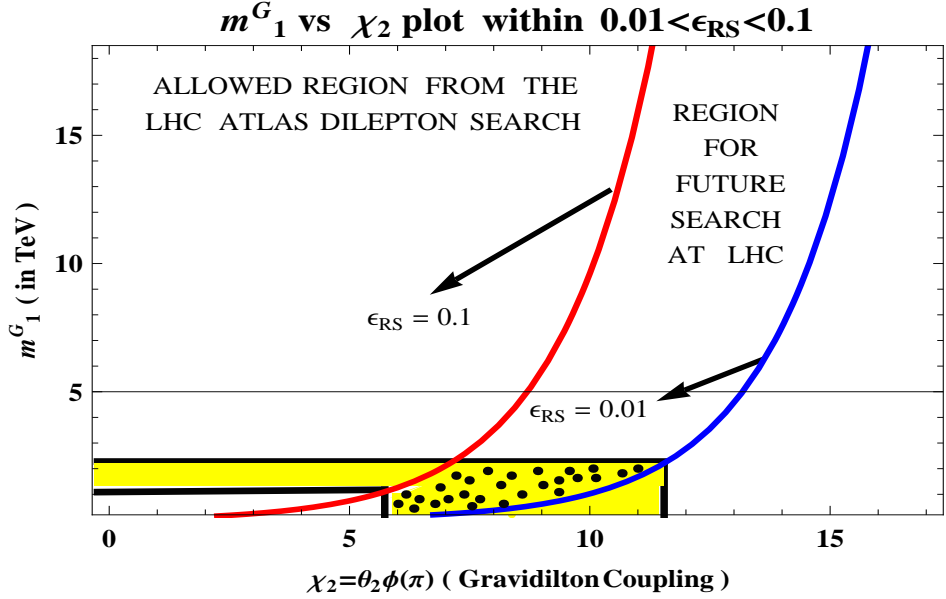}
\caption{Variation of the lightest Kaluza-Klein graviton
 mass $m^{G}_{1}$ for $n=1$ mode with respect to the dilaton coupling $\chi_{2}=\theta_{2}\phi(\pi)$ for the proposed theoretical setup for
 $0.01<\epsilon_{RS}<0.10$ at the wall of the TeV brane. We have also shown the present status
of the allowed region for the lower limit of the lightest Kaluza-Klein graviton
 mass for LHC ATLAS dilepton search by the yellow shaded region bounded by black coloured line drawn 
for $\epsilon_{RS}=0.10$ and $\epsilon_{RS}=0.01$ respectively.
 Here for this plot we fix, $A_{1}=0.05$, $\theta_{2}=-\theta_{1}$ and $\alpha_{5}\sim 5\times 10^{-7}$. Additionally, the white region bounded by the red and blue curve represents the future probing 
region for LHC. Also the black dotted region represents the overlapping area between the parameter space obtained from the 
prposed model and present LHC ATLAS dilepton search.}
\label{f3}
\end{figure*}

\begin{figure*}[htb]
\centering
\subfigure[$m^{G}_{1}$~vs~$\alpha_{5}$~plot~for~$\epsilon_{RS}=0.01$]{
    \includegraphics[width=8.5cm,height=8cm] {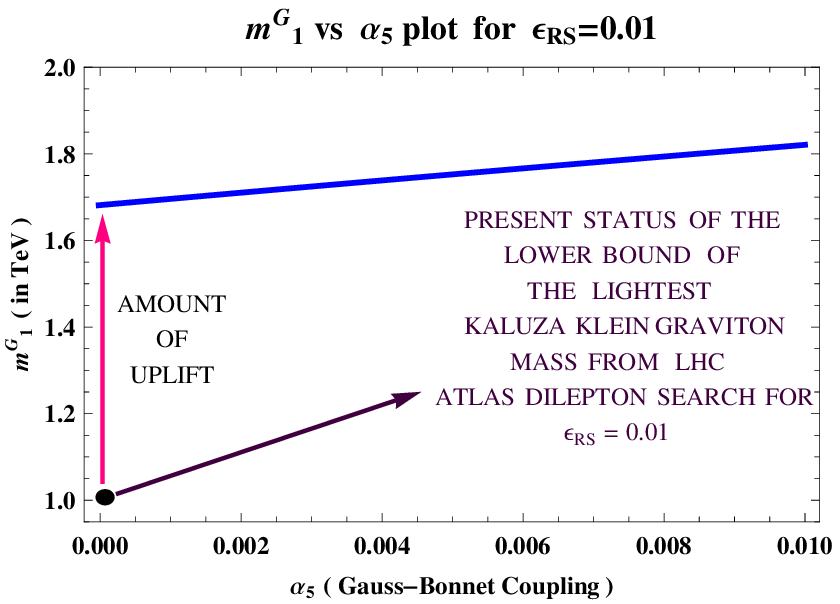}
    \label{figs1}
}
\subfigure[$m^{G}_{1}$~vs~$\alpha_{5}$~plot~for~$\epsilon_{RS}=0.1$]{
    \includegraphics[width=8.5cm,height=8cm] {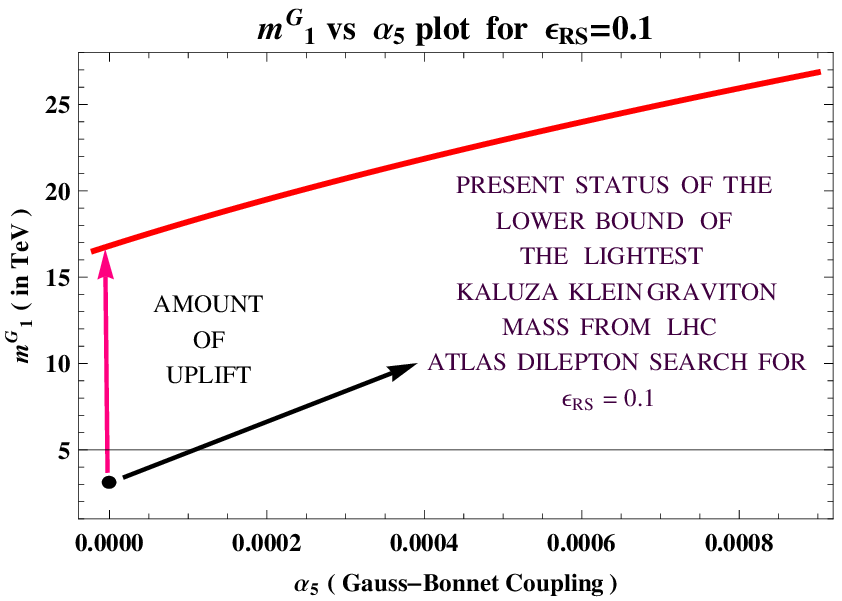}
    \label{figs2}
}
\caption[Optional caption for list of figures]{Variation of the lightest Kaluza-Klein graviton
 mass $m^{G}_{1}$ with respect to 5D Gauss-Bonnet coupling $\alpha_{5}$ for $n=1$ mode with \subref{figs1} 
$\epsilon_{RS}=0.01$ and \subref{figs2} 
$\epsilon_{RS}=0.1$ for the proposed theoretical setup at the wall of the TeV brane. We have also shown the present status
of the lower limit of the lightest Kaluza-Klein graviton
 mass for LHC ATLAS dilepton search by the black coloured point drawn 
for $\epsilon_{RS}=0.10$ and $\epsilon_{RS}=0.01$ respectively.
 Here for this plot we fix, $A_{1}=0.05$, $\theta_{2}=-\theta_{1}$, $\theta_{2}\phi(\pi)\sim 11$ and $\theta_{1}\phi(\pi)\sim -11$. 
Additionally, we have shown the amount of the uplift of the lower bound of the lightest Kaluza-Klein graviton mass compared to the result
obtained from the LHC ATLAS dilepton search.
}
\label{f2}
\end{figure*}

Also the localized brane tension can be computed as:
\begin{widetext}
\be\begin{array}{llll}\label{brten}
    \displaystyle V^{\pm}_{2}=-V^{\mp}_{1}=\pm 24 k_{\pm}M^{3}_{(5)}e^{-\theta_{2}\phi(y)}\left[1-\frac{\alpha_{5}
\left(1-A_{1}e^{\theta_{1}\phi(y)}\right)}{3M^{2}_{5}}k^{2}_{\pm}r^{2}_{c}\right].
   \end{array}\ee
\end{widetext}
where the brane tension $V_{1}$ and $V_{2}$ are localized at the position of orbifold fixed points, $y_{i}=[0,\pi]$, where
the hidden and visible branes are placed respectively. However, it is clearly observed from Eq~(\ref{eq2}) and Eq~(\ref{brten}) that within the bulk 
both the warp factor and the brane tension varies with the extra dimensional coordinate 'y' due to the presence of the dynamical dilaton degrees 
of freedom within the bulk.
Here we have discarded 
the other branch of solution of $k_{\bf +}$ (+ve branch) which diverges in the limit $\alpha_{5}\rightarrow 0$, bringing 
in ghost fields \cite{Rizzo:2004rq,Dotti:2007az,Torii:2005xu,Konoplya:2010vz,Kim:2000ym,Nojiri:2010wj}. So that we are concentrating 
only on the -ve branch of the solution which we call further as, $k_{-}:=k_{\bf M}$. 

 In the limit $\alpha_{5}\rightarrow 0$, $A_{1}\rightarrow 0$, $\theta_{1}\rightarrow 0$ and  $\theta_{2}\rightarrow 0$ we
 retrieve asymptotically the same result as in the case of RS model with \cite{Randall:1999ee,Randall:1999vf}:
\be k_{\bf M}\rightarrow k_{RS}=\sqrt{-\frac{\Lambda_{5}}{24M^{3}_{5}}}\ee  and the barne tension is given by,
 \be V^{-}_{2}\rightarrow V^{RS}_{2}=24M^{3}_{5}k_{RS}\ee with  $\Lambda_{5}<0$.

 Now expanding Eq~(\ref{eq2}) in the perturbation series order by order around 
$\alpha_{5}\rightarrow 0$, $A_{1}\rightarrow 0$, $\theta_{1}\rightarrow 0$ and  $\theta_{2}\rightarrow 0$  we can write:
\begin{widetext}
 \be\begin{array}{lll}\label{jkq1}
\displaystyle k_{\bf M}=k_{RS}~e^{\frac{\theta_{2}\phi(y)}{2}}\sqrt{\left[1+\frac{8\alpha_{5} k^{2}_{RS}}{M^{2}_{5}(1-A_{1}e^{\theta_{1}\phi(y)})}
\left\{1-2e^{(\theta_{1}+\theta_{2})\phi(y)}A_{1}+e^{(2\theta_{1}+\theta_{2})\phi(y)}A^{2}_{1}+\cdots\right\}
+{\cal O}\left(\frac{\alpha^{2}_{5}k^{4}_{RS}}{M^{4}_{5}}\right)+\cdots\right]}.
\end{array}\ee
\end{widetext}

For the graviton, the Kaluza Klein mass spectra for n-th excited state in presence of gavidilatonic and Gauss-Bonnet coupling by applying Neumann (-) and Dirichlet (+)
 boundary conditions at the orbifold point $y_{i}=\pi$ where the visible brane is placed,
can be written as \cite{Choudhury:2013yg,Gherghetta:2010cj}:
\be\begin{array}{lll}\label{wq1}
    \displaystyle m^{\bf G}_{n}=\left(n+\frac{1}{2}\mp \frac{1}{4}\right)\pi k_{\bf M}(\pi)~e^{-k_{\bf M}r_{c}\pi}
   \end{array}\ee
 in presence Gauss-Bonnet coupling and string loop correction. For the numerical estimations we use the +ve Dirichlet branch throughout the article.
 Furthermore the modified 4D effective Planck mass in presence of Gauss-Bonnet coupling can be expressed in terms of 5D mass scale as\cite{Choudhury:2013yg}:
\be\begin{array}{llll}\label{wq2}
    \displaystyle M^{2}_{Pl}=\frac{M^{3}_{5}}{k_{\bf M}}\left(1-e^{-2k_{\bf M}r_{c}\pi}\right).
   \end{array}\ee
Now using Eq~(\ref{jkq1}) on Eq~(\ref{wq2}) the 5D quantum garvity scale at the position of visible brane $y=\pi$ turns out to be:
\begin{widetext}
\be\begin{array}{lll}\label{fived}
    \displaystyle M_{5}=\sqrt[3]{Z_{\bf T}}M_{Pl}~e^{\frac{\theta_{2}\phi(\pi)}{6}}\left[1+\frac{8\alpha_{5}Z^{\frac{4}{3}}_{\bf T}e^{-\frac{\theta_{2}\phi(\pi)}{3}}}{(1-A_{1}e^{\theta_{1}\phi(\pi)})}
\left\{1-2e^{(\theta_{1}+\theta_{2})\phi(\pi)}A_{1}+e^{(2\theta_{1}+\theta_{2})\phi(\pi)}A^{2}_{1}+\cdots\right\}
\right.\\ \left.\displaystyle~~~~~~~~~~~~~~~~~~~~~~~~~~~~~~~~~~~~~~~~~~~~~~~~~~~~~~~~~~~~~~~~~~~~~~~~~~~~~~~~~~~~~~~~~~~~~~~~~~~~~~~~~
+{\cal O}\left(\alpha^{2}_{5}Z^{\frac{8}{3}}_{\bf T}e^{-\frac{2\theta_{2}\phi(\pi)}{3}}\right)+\cdots\right]^{\frac{1}{6}}
\end{array}\ee
\end{widetext}
 where we use the fact that, $e^{-2k_{\bf M}r_{c}\pi}<<1$ approximation holds good in Eq~(\ref{wq2}).
Here additionally we use the fact that, $k_{RS}=Z_{\bf T}M_{PL}$, where $Z_{\bf T}$ is a dimensionless tuning parameter.
Now for the sake of clarity one can write $Z_{\bf T}$ as:
\be\begin{array}{lll}\label{ghrt}
    \displaystyle Z_{\bf T}=\frac{M_{5}}{M_{Pl}}\epsilon_{RS}
   \end{array}\ee
where we introduce an additional parameter, $\epsilon_{RS}=\frac{k_{RS}}{M_{5}}$
with the restriction on the parameter, $0.01<\epsilon_{RS}<0.1$, as used in \cite{ATLAS:2011ab,Das:2013lqa}. This requirement
emerges from the fact that the bulk curvature must be smaller
than the Planck scale so that the classical solutions for the bulk metric given
by the proposed model can be trusted. On the other hand string theory
also supports this favoured range within the background of Klebanov Strassler throat geometry motivated $D3-\overline{D3}$ brane-antibrane setup \cite{Davoudiasl:1999jd}.
It is important to mention here that only for RS model $Z_{\bf T}\approx \epsilon_{RS}$, as $M_{5}\sim M_{Pl}$ approximation holds good in RS setup. 
Further substituting Eq~(\ref{ghrt}) in Eq~(\ref{fived}) we found the simplified expression for the 5D quantum gravity scale in terms of $\epsilon_{RS}$ which 
turns out to be:
\begin{widetext}
\be\begin{array}{lll}\label{fivedx}
    \displaystyle M_{5}=\sqrt{\epsilon_{RS}}M_{Pl}~e^{\frac{\theta_{2}\phi(\pi)}{4}}\left[1+\frac{8\alpha_{5}\epsilon^{2}_{RS}}{(1-A_{1}e^{\theta_{1}\phi(\pi)})}
\left\{1-2e^{(\theta_{1}+\theta_{2})\phi(\pi)}A_{1}+e^{(2\theta_{1}+\theta_{2})\phi(\pi)}A^{2}_{1}+\cdots\right\}
+{\cal O}\left(\alpha^{2}_{5}\epsilon^{4}_{RS}\right)+\cdots\right]^{\frac{1}{4}}

   \end{array}\ee
\end{widetext}

On the other hand to solve the hierarchy problem the brane localized Higgs mass can be written as: 
\be\begin{array}{llll}\label{massasdphi}
   \displaystyle  m_{H}\approx
m_{\bf CUT}~e^{-k_{\bf M} r_{c}\pi}
   \end{array}\ee
where we introduce a new parameter $m_{\bf CUT}$ defined as,
  \be \label{qg}
m_{\bf CUT}= M_{Pl}\ee
 physically represents the cut off scale of the theory, above which new physics beyond standard model is expected
to appear. A natural choice for this would be Planck or quantum gravity scale
beyond which standard model will not be valid. 

Now using Eq~(\ref{jkq1}) we introduce a new parameter $\epsilon_{\bf M}$ defined as:
\begin{widetext}
 \be\begin{array}{lll}\label{jkq2}
\displaystyle \epsilon_{\bf M}=\frac{k_{\bf M}}{M_{5}}\approx Z^{\frac{2}{3}}_{\bf T}~e^{\frac{\theta_{2}\phi(\pi)}{3}}
\left[1+\frac{8\alpha_{5}Z^{\frac{4}{3}}_{\bf T}e^{-\frac{\theta_{2}\phi(\pi)}{3}}}{(1-A_{1}e^{\theta_{1}\phi(\pi)})}
\left\{1-2e^{(\theta_{1}+\theta_{2})\phi(\pi)}A_{1}+e^{(2\theta_{1}+\theta_{2})\phi(\pi)}A^{2}_{1}+\cdots\right\}
\right.\\ \left.\displaystyle~~~~~~~~~~~~~~~~~~~~~~~~~~~~~~~~~~~~~~~~~~~~~~~
~~~~~~~~~~~~~~~~~~~~~~~~~~~~~~~~~~~~~~~~~~~~~~~~~~~~~~~~
+{\cal O}\left(\alpha^{2}_{5}Z^{\frac{8}{3}}_{\bf T}e^{-\frac{2\theta_{2}\phi(\pi)}{3}}\right)+\cdots\right]^{\frac{1}{3}}\\
\displaystyle~~~=\epsilon_{RS}~e^{\frac{\theta_{2}\phi(\pi)}{2}}\left[1+\frac{8\alpha_{5} \epsilon^{2}_{RS}}{(1-A_{1}e^{\theta_{1}\phi(\pi)})}
\left\{1-2e^{(\theta_{1}+\theta_{2})\phi(\pi)}A_{1}+e^{(2\theta_{1}+\theta_{2})\phi(\pi)}A^{2}_{1}+\cdots\right\}
+{\cal O}\left(\alpha^{2}_{5}\epsilon^{4}_{RS}\right)+\cdots\right]^{\frac{1}{2}}.
\end{array}\ee
\end{widetext}
 Further using Eq~(\ref{jkq2}) and Eq~(\ref{massasdphi}) in the graviton Kaluza Klein mass spectra as stated in Eq~(\ref{wq1}), 
the first Kaluza-Klein excitation ($n=1$) 
becomes:
\begin{widetext}
 \be\begin{array}{lll}\label{jkq3z}
\displaystyle m^{\bf G}_{1}=x_{1}\epsilon_{\bf M}m_{H}\left(1-e^{-2k_{\bf M}r_{c}\pi}\right)^{\frac{1}{3}}
\\\displaystyle~~~~~\approx x_{1} Z^{\frac{2}{3}}_{\bf T}m_{H}~e^{\frac{\theta_{2}\phi(\pi)}{3}}
\left[1+\frac{8\alpha_{5}Z^{\frac{4}{3}}_{\bf T}e^{-\frac{\theta_{2}\phi(\pi)}{3}}}{(1-A_{1}e^{\theta_{1}\phi(\pi)})}
\left\{1-2e^{(\theta_{1}+\theta_{2})\phi(\pi)}A_{1}+e^{(2\theta_{1}+\theta_{2})\phi(\pi)}A^{2}_{1}+\cdots\right\}
\right.\\ \left.\displaystyle~~~~~~~~~~~~~~~~~~~~~~~~~~~~~~~~~~~~~~~~~~~~~~~
~~~~~~~~~~~~~~~~~~~~~~~~~~~~~~~~~~~~~~~~~~~~~~~~~~~~~~~~~~~~~~~~~~
+{\cal O}\left(\alpha^{2}_{5}Z^{\frac{8}{3}}_{\bf T}e^{\frac{2\theta_{2}\phi(\pi)}{3}}\right)+\cdots\right]^{\frac{1}{3}}
\\\displaystyle~~~~~\approx x_{1} \epsilon_{RS} m_{H}~e^{\frac{\theta_{2}\phi(\pi)}{2}}
\left[1+\frac{8\alpha_{5}\epsilon^{2}_{RS}}{(1-A_{1}e^{\theta_{1}\phi(\pi)})}
\left\{1-2e^{(\theta_{1}+\theta_{2})\phi(\pi)}A_{1}+e^{(2\theta_{1}+\theta_{2})\phi(\pi)}A^{2}_{1}+\cdots\right\}
+{\cal O}\left(\alpha^{2}_{5}\epsilon^{4}_{RS}\right)+\cdots\right]^{\frac{1}{2}}
\\\displaystyle~~~~~\approx (m^{\bf G}_{1})_{\bf RS}{\bf \Theta}_{\bf T}
\end{array}\ee
\end{widetext}
where we again use the fact that, $e^{-2k_{\bf M}r_{c}\pi}<<1$ and the lightest graviton mass for Randall Sundrum model 
is given by:
\be\label{mass}(m^{\bf G}_{1})_{\bf RS}=x_{1}\epsilon^{\frac{2}{3}}_{RS}m_{H}\ee
where $x_{1}=7\pi/4$ be the root of the Beesel function of the order 1 as obtained from Eq~(\ref{wq1}).
 Here in Eq~(\ref{jkq3z}) we introduce a new parameter, ${\bf \Theta}_{\bf T}$ given by:
\begin{widetext}
 \be\begin{array}{lll}\label{th1}
 \displaystyle    {\bf \Theta}_{\bf T}=\epsilon^{\frac{1}{3}}_{RS}~e^{\frac{\theta_{2}\phi(\pi)}{2}}
\left[1+\frac{8\alpha_{5}\epsilon^{2}_{RS}}{(1-A_{1}e^{\theta_{1}\phi(\pi)})}
\left\{1-2e^{(\theta_{1}+\theta_{2})\phi(\pi)}A_{1}+e^{(2\theta_{1}+\theta_{2})\phi(\pi)}A^{2}_{1}+\cdots\right\}
+{\cal O}\left(\alpha^{2}_{5}\epsilon^{4}_{RS}\right)+\cdots\right]^{\frac{1}{2}}
    \end{array}\ee
\end{widetext}
which signifies the {\it multiplicative uplifting
factor} of the lightest Kaluza Klein graviton mass spectra for the proposed model compared to the lightest graviton mass for Randall Sundrum model.

The five dimensional action describing the interaction between bulk graviton and visible Standard Model
fields dominated by fermionic contribution on the brane is given by
\be\begin{array}{llll}\label{delq1}
 \displaystyle {\cal S}_{\bf SM-\bf G}=-\frac{{\cal K}_{(5)}}{2}\int d^{5}x\sqrt{-g_{(5)}}{\bf T}^{\alpha\beta}_{\bf SM}(x){\bf h}_{\alpha\beta}(x,y)\delta(y-\pi)   
   \end{array}\ee
where ${\bf T}^{\alpha\beta}_{\bf SM}(x)$ represents the energy momentum or stress energy tensor containing all informations of Standard Model matter fields 
on the visible brane and $h_{\alpha\beta}(x,y)$ be the bulk graviton degrees of freedom. In this context ${\cal K}_{(5)}:=\frac{2}{M^{\frac{3}{2}}_{(5)}}$ is the coupling strength describing the tensor fluctuation in the context of graviton phenomenology.
After substituting the Kaluza-Klien expansion for graviton degrees of freedom:
\be\begin{array}{lllll}\label{KK41g}
   \displaystyle {\bf h}_{\alpha\beta}(x,y)=\sum^{\infty}_{n=0}{\bf h}^{(n)}_{\alpha\beta}(x)~\frac{\chi^{(n)}_{\bf G}(y)}{\sqrt{r_{c}}}.
   \end{array}\ee
and rescaling the fields appropriately, the effective four dimensional action turns out to be
\begin{widetext}\be\begin{array}{llll}\label{delq2}
 \displaystyle {\cal S}_{\bf SM-\bf G}=-\frac{{\cal K}_{(5)}}{2}\int d^{5}x~r_{c}~e^{-4A(y)}
{\bf T}^{\alpha\beta}_{\bf SM}(x)\sum^{\infty}_{n=0}{\bf h}^{(n)}_{\alpha\beta}(x)\frac{\chi^{(n)}_{\bf G}(y)}{\sqrt{r_{c}}}\delta(y-\pi)\\
\displaystyle ~~~~~~~~~~=-\frac{\sqrt{r_{c}}{\cal K}_{(5)}}{2}\int d^{4}x~e^{-4A(\pi)}
{\bf T}^{\alpha\beta}_{\bf SM}(x)\sum^{\infty}_{n=0}{\bf h}^{(n)}_{\alpha\beta}(x)\chi^{(n)}_{\bf G}(\pi)\\
\displaystyle ~~~~~~~~~~=-\frac{\sqrt{k_{\bf M}}r_{c}{\cal K}_{(5)}}{2}\int d^{4}x~
{\bf T}^{\alpha\beta}_{\bf SM}(x)\left[{\bf h}^{(0)}_{\alpha\beta}(x)
+ e^{k_{\bf M}r_{c}\pi}\sum^{\infty}_{n=1}{\bf h}^{(n)}_{\alpha\beta}(x)\right]\\
\displaystyle ~~~~~~~~~~=-\frac{r_{c}}{k_{RS}M_{Pl}}~e^{-\frac{\theta_{2}\phi(\pi)}{2}}
\left[1+\frac{8\alpha_{5} k^{2}_{RS}}{M^{2}_{5}(1-A_{1}e^{\theta_{1}\phi(\pi)})}
\left\{1-2e^{(\theta_{1}+\theta_{2})\phi(\pi)}A_{1}+e^{(2\theta_{1}+\theta_{2})\phi(\pi)}A^{2}_{1}+\cdots\right\}
\right.\\ \left.  \displaystyle ~~~~~~~~~~~~~~~~~~~~~~~~~~~~~~~~~~~~~~~~~~~~~~+{\cal O}\left(\frac{\alpha^{2}_{5}
k^{4}_{RS}}{M^{4}_{5}}\right)+\cdots\right]^{-\frac{1}{2}}\displaystyle \int d^{4}x~
{\bf T}^{\alpha\beta}_{\bf SM}(x)\left[{\bf h}^{(0)}_{\alpha\beta}(x)
+ e^{k_{\bf M}r_{c}\pi}\sum^{\infty}_{n=1}{\bf h}^{(n)}_{\alpha\beta}(x)\right].   
   \end{array}\ee\end{widetext}
It is evident from equation(\ref{delq2}) that while the zero mode couples to the brane fields with usual gravitational coupling $\sim 1/M_{Pl}$
 which we have taken as unity,
the coupling of the KK modes are $\sim  e^{k_{\bf M}r_{c}\pi} / M_{Pl} \sim ~ TeV^{-1}$ which is much larger than the coupling of massless graviton.
Though such feature is also observed for the graviton KK modes in the usual RS model, here the graviton KK mode coupling depends on the GB coupling $\alpha_{5}$.
 In the present context the values of  $k_{\bf M}$ though increases with $\alpha_{5}$, the enhancement of the graviton KK mode mass causes the overall decrease
in the detection cross section. Thus the absence of any signature of graviton KK modes, as reported by ATLAS data in dilepton decay processes, may be
explained by GB coupling in warped geometry models. 
\begin{figure*}[htb]
\centering
\subfigure[$m^{G}_{1}$~vs~$A_{1}$~plot~for~$\epsilon_{RS}=0.01$]{
    \includegraphics[width=8.5cm,height=8cm] {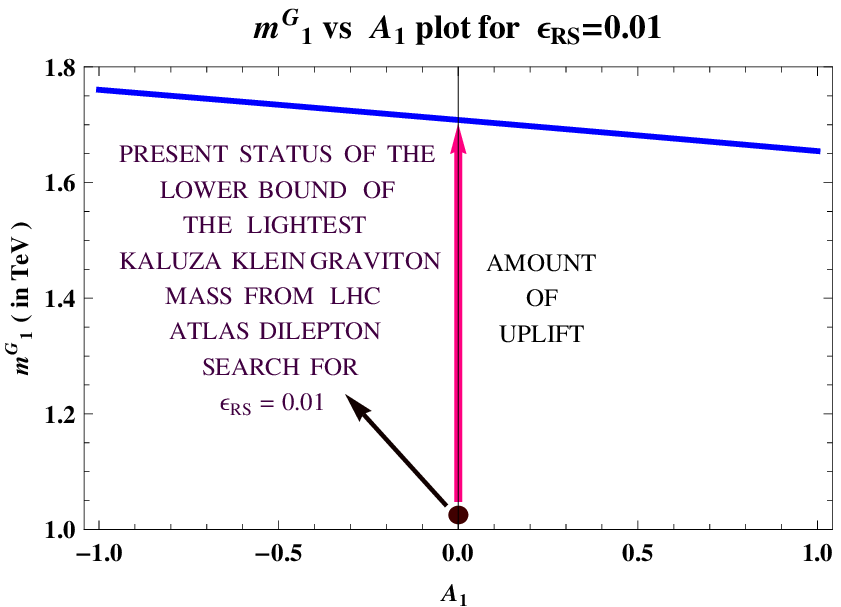}
    \label{figs11}
}
\subfigure[$m^{G}_{1}$~vs~$A_{1}$~plot~for~$\epsilon_{RS}=0.1$]{
    \includegraphics[width=8.5cm,height=8cm] {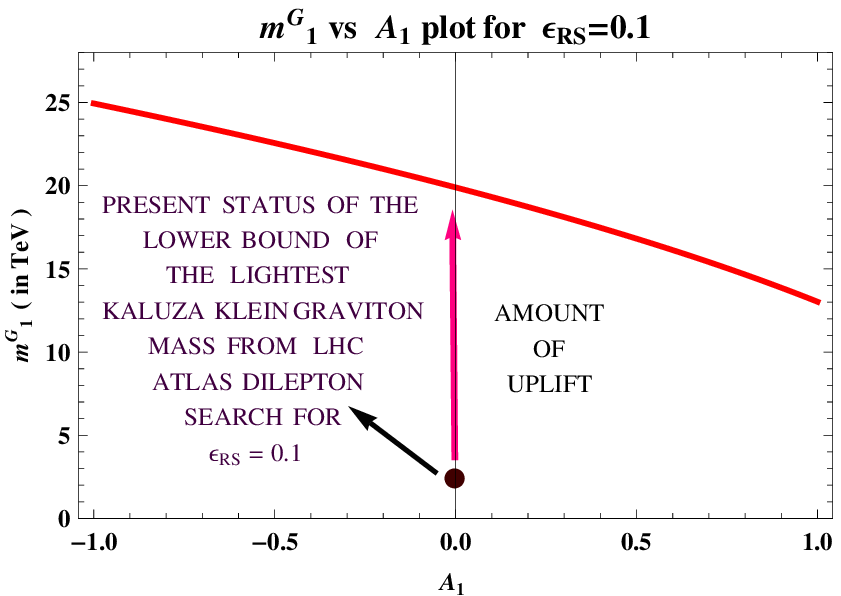}
    \label{figs22}
}
\caption[Optional caption for list of figures]{Variation of the lightest Kaluza-Klein graviton
 mass $m^{G}_{1}$ with respect to string two-loop coupling $A_{1}$ for $n=1$ mode with \subref{figs11} 
$\epsilon_{RS}=0.01$ and \subref{figs22} 
$\epsilon_{RS}=0.1$ for the proposed theoretical setup at the wall of the TeV brane. We have also shown the present status
of the lower limit of the lightest Kaluza-Klein graviton
 mass for LHC ATLAS dilepton search by the black coloured point drawn 
for $\epsilon_{RS}=0.10$ and $\epsilon_{RS}=0.01$ respectively.
 Here for this plot we fix,  $\theta_{2}=-\theta_{1}$, $\theta_{2}\phi(\pi)\sim 11$ and $\theta_{1}\phi(\pi)\sim -11$. 
Additionally, we have shown the amount of the uplift of the lower bound of the lightest Kaluza-Klein graviton mass compared to the result
obtained from the LHC ATLAS dilepton search.
}
\label{f4}
\end{figure*}

In table~(\ref{tab1}) we present a comparative study between the lower limit of the lightest Kaluza-Klein graviton
 mass for $n=1$ mode from the proposed theoretical model, the well known
Randall-Sundrum (RS) model and the LHC ATLAS dilepton search in 7 TeV
proton-proton collision. Additionally we have shown that the 5D mass scale of the proposed model is lying within the window 
$1.56M_{Pl}<M_{5}<4.95M_{Pl}$ for $0.01<\epsilon_{RS}<0.1$. 
For RS model, the lower limit of the graviton KK mode mass lying within the window,
 $0.22 ~TeV<m^{G,{\bf RS}}_{1}<1.02~TeV $ for $0.01<\epsilon_{RS}<0.1$. On the other hand the latest data from ATLAS 
predicts the graviton KK mode mass lying within $1.01 ~TeV<m^{G,{\bf ATLAS}}_{1}<2.22~TeV$ for $0.01<\epsilon_{RS}<0.1$. This implies a serious conflict
between graviton KK modes as predicted in RS model and the result reported
by ATLAS Collaboration. But for the proposed model the lightest bound of the KK graviton mass estimated 
as $1.68 ~TeV<m^{G,{\bf RS}}_{1}<16.82~TeV $ which is above the recent lower bound of the KK graviton mass measured from LHC ATLAS dilepton search
and lies within the parameter space for the future probing region of LHC. By taking into consideration of the enhancement of coupling between SM fields and graviton,
it may be observed from the table~(\ref{tab1}) that for $\epsilon_{RS}=0.07$ or higher, the lower bound of the graviton KK mode exceeds that predicted from ATLAS data.

To study the various hidden phenomenological features within super-Planckian regime of the UV cut-off scale from our proposed setup the scanned parameter space 
is given by:
\be\begin{array}{llll}\label{fg1}
    \displaystyle ~~~~\alpha_{5}={\cal O}((4.8-5.1)\times 10^{-7}),\\
\displaystyle ~~~~|A_{1}|\sim {\cal O}(0.01-0.09),\\
\displaystyle ~~~~\theta_{2}=-\theta_{1},\\
\displaystyle ~~~~\chi_{1}={\theta_{1}\phi(\pi)}\sim -11,\\
\displaystyle ~~~~\chi_{2}={\theta_{2}\phi(\pi)}\sim 11, \\
\displaystyle for~~m_{H}\sim {\cal O}((125-126)~GeV). 
   \end{array}\ee
This bound on GB coupling $\alpha_{5}$ is also consistent with the solar system constraint \cite{Chakraborty:2012sd}, combined constraint from the
 Higgs mass and favoured decay channels $H\rightarrow (\gamma\gamma,\tau\bar{\tau})$ \cite{ATLAS:2013mma} using ATLAS \cite{Aad:2012tfa} and CMS \cite{Chatrchyan:2013lba} data
within the $5\sigma$ statistical C.L. Additionally, the parameter space mentioned in Eq~(\ref{fg1}) are 
necessary ingredient to increase/uplift the lower bound of the lightest KK graviton mass constrained from LHC ATLAS dilepton search.
It is important to mention here that the bound on the 5D Gauss-Bonnet coupling 
obtained from Eq~(\ref{fg1}) is lying below the upper cut-off on coupling 
obtained from the Kubo formula i.e. $\alpha_{5}<1/4$ \cite{Choudhury:2013eoa,Choudhury:2013dia,Brigante:2007nu}
 obtained in the context of ${\bf AdS_{5}/CFT_{4}}$ correspondence.

 In Fig~(\ref{f1}) we have shown the variation of the lightest Kaluza-Klein graviton
 mass from the proposed model (represented by red curve) and Randall Sundrum (RS) model (represented by blue curve) with respect to 
the phenomenological parameter {\large$ \epsilon_{RS}=\frac{k_{RS}}{M_{5}}$}, within the range $0.01<\epsilon_{RS}<0.10$ as
stated in Eq~(\ref{fg1}). We have also shown the present status
of the lower limit of the lightest Kaluza-Klein graviton
 mass for LHC ATLAS dilepton search by the green curve in Fig~(\ref{f1}).
Here the allowed region is in the upper half  
of the green curve. The rest of the region below the geen curve phenomenologically is ruled out.
 We have also explored the phenomenological feature of the lightest Kaluza-Klein graviton
 mass with respect to the dialton coupling $\chi_{2}=\theta_{2}\phi(\pi)$ with the 5D ${\bf AdS_{5}}$ cosmological constant $\Lambda_{5}$
at the wall of the visible brane for the proposed theoretical setup in Fig~(\ref{f3}). We have also shown the present status
of the allowed region for the lower limit of the lightest Kaluza-Klein graviton
 mass for LHC ATLAS dilepton search by the yellow shaded region in Fig~(\ref{f3}). This will constrain the parameter $\chi_{2}$ within,
$\chi_{2}\sim{\cal O}(6-12.8)$. This is also consistent with the present theoretical analysis as the proposed setup predicts $\chi_{2}\sim 11$ 
as mentioned in Eq~(\ref{fg1}). For both the branch the lightest Kaluza-Klein graviton
 mass increases exponentially by increasing the dialton coupling $\chi_{2}=\theta_{2}\phi(\pi)$ and fixing the other parameters within the allowed parameter 
space stated in Eq~(\ref{fg1}).
Next in Fig~(\ref{figs1}) and Fig~(\ref{figs2}) we have presented the characteristic feature of the lightest Kaluza-Klein graviton
 mass with respect to the 5D Gauss-Bonnet coupling ($\alpha_{5}$) for the proposed theoretical setup for
$\epsilon_{RS}=0.01$ and $\epsilon_{RS}=0.10$ respectively. For both the cases the lightest Kaluza-Klein graviton
 mass increases by increasing the 5D Gauss-Bonnet coupling ($\alpha_{5}$) and fixing the other parameters stated in Eq~(\ref{fg1}). We have also shown the present status
of the lower limit of the lightest Kaluza-Klein graviton
 mass for LHC ATLAS dilepton search by a point in Fig~(\ref{figs1}) and Fig~(\ref{figs2}) both. To uplift/increase the lower bound of the 
lightest Kaluza-Klein graviton
 mass estimated from the proposed theoretical setup compared to the LHC dilepton search by proposing the 
5D Gauss-Bonnet coupling ($\alpha_{5}$) within , $\alpha_{5}\sim {\cal O}((4.8-5.1)\times 10^{-7})$, as explicitly mentioned in the table~(\ref{tab1}).
Finally, in Fig~(\ref{figs11}) and Fig~(\ref{figs22}) we have explicitly shown the behaviour of the lightest Kaluza-Klein graviton
 mass with respect to the string two-loop coupling $A_{1}$ by fixing the rest of the parameters for the proposed theoretical setup for
$\epsilon_{RS}=0.01$ and $\epsilon_{RS}=0.10$ respectively. For both the cases the lightest Kaluza-Klein graviton
 mass decreases by increasing the string two-loop coupling $A_{1}$ and fixing the other parameters stated in Eq~(\ref{fg1}). We have also shown the present status
of the lower limit of the lightest Kaluza-Klein graviton
 mass for LHC ATLAS dilepton search by a point in these figures.

To summarize, we say that the perturbative two loop higher genus correction to Einstein’s gravity in presence of stringy type IIB 
gravidilatonic interaction can also be
examined through collider experimental tests by studying the hidden phenomenological features of lightest KK mode from graviton mass spectrum. Using the
prescription mentioned in this paper one can directly check the
validity and justifiability of a higher order gravity or any modified gravity
model in presence of stringy higher genus corrections and also constrain the associated parameter space which involves various couplings with
such higher order gravity corrections. Thus, in this work,
by applying the requirements from latest data
we have also elaborately analyzed the multi parameter space dependence on the lightest Kaluza-Klein graviton mass by studying the flow of the running through
the crucial parameters proposed in this article. This analysis
therefore determines the allowed parameter space for the proposed model and
brings out the phenomenological constraint on the value
of the stringy parameters in the context of recent LHC experiment by scanning the multiparameter space within a phenomenologically feasible range.


\section*{Acknowledgments}
SC thanks Council of Scientific and
Industrial Research, India for financial support through Senior
Research Fellowship (Grant No. 09/093(0132)/2010). SC also thanks The Abdus Salam International Center for Theoretical Physics,Trieste, Italy and the 
organizers of SUSY 2013 conference for the hospitality during the work.





\begin{references}

\bibitem{Choudhury:2012yh}
  S.~Choudhury and S.~Pal,
  Nucl.\ Phys.\ B {\bf 874} (2013) 85
  [arXiv:1208.4433 [hep-th]].

\bibitem{Choudhury:2012kw}
  S.~Choudhury and S.~Pal,
  arXiv:1210.4478 [hep-th].

\bibitem{Choudhury:2013qza}
  S.~Choudhury and A.~Dasgupta,
  Nucl.\ Phys.\ B {\bf 882} (2014) 195
  [arXiv:1309.1934 [hep-ph]].

\bibitem{Choudhury:2013eoa}
  S.~Choudhury, S.~Sadhukhan and S.~SenGupta,
  arXiv:1308.1477 [hep-ph].

\bibitem{Choudhury:2013dia}
  S.~Choudhury and S.~SenGupta,
  arXiv:1306.0492 [hep-th].

\bibitem{Choudhury:2013yg}
  S.~Choudhury and S.~SenGupta,
  JHEP {\bf 1302} (2013) 136
  [arXiv:1301.0918 [hep-th]].

\bibitem{Choudhury:2014hna}
  S.~Choudhury, J.~Mitra and S.~SenGupta,
  JHEP {\bf 1408} (2014) 004
  [arXiv:1405.6826 [hep-th]].

\bibitem{Kim:1999dq}
  J.~E.~Kim, B.~Kyae and H.~M.~Lee,
  Phys.\ Rev.\ D {\bf 62} (2000) 045013
  [hep-ph/9912344].

\bibitem{Lee:2000vf}
  H.~M.~Lee,
  hep-th/0010193.

\bibitem{Kim:2000pz}
  J.~E.~Kim, B.~Kyae and H.~M.~Lee,
  Nucl.\ Phys.\ B {\bf 582} (2000) 296
   [Erratum-ibid.\ B {\bf 591} (2000) 587]
  [hep-th/0004005].

\bibitem{Kim:2000ym}
  J.~E.~Kim and H.~M.~Lee,
  Nucl.\ Phys.\ B {\bf 602} (2001) 346
   [Erratum-ibid.\ B {\bf 619} (2001) 763]
  [hep-th/0010093].

\bibitem{Green1}
  Superstring Theory. Vol. 1: Introduction - Green, Michael B. et al. Cambridge, Uk: Univ. Pr. ( 1987) 469 P. ( Cambridge Monographs On Mathematical Physics).

\bibitem{Green2} 
 Superstring Theory. Vol. 2: Loop Amplitudes,
 Anomalies And Phenomenology - Green, Michael B. et al. Cambridge, Uk: Univ. Pr. ( 1987) 596 P. ( Cambridge Monographs On Mathematical Physics).

\bibitem{Polchinski1}
String theory. Vol. 1: An introduction to the bosonic string - Polchinski, J. Cambridge, UK: Univ. Pr. (1998) 402 p.

\bibitem{Polchinski2}
String theory. Vol. 2: Superstring theory and beyond - Polchinski, J. Cambridge, UK: Univ. Pr. (1998) 531 p.

\bibitem{Randall:1999ee}
  L.~Randall and R.~Sundrum,
  Phys.\ Rev.\ Lett.\  {\bf 83} (1999) 3370
  [hep-ph/9905221].

\bibitem{Randall:1999vf}
  L.~Randall and R.~Sundrum,
  Phys.\ Rev.\ Lett.\  {\bf 83} (1999) 4690
  [hep-th/9906064].

\bibitem{Goldberger:1999uk}
  W.~D.~Goldberger and M.~B.~Wise,
  Phys.\ Rev.\ Lett.\  {\bf 83} (1999) 4922
  [hep-ph/9907447].


\bibitem{Goldberger:1999un}
  W.~D.~Goldberger and M.~B.~Wise,
  Phys.\ Lett.\ B {\bf 475} (2000) 275
  [hep-ph/9911457].

\bibitem{Goldberger:1999wh}
  W.~D.~Goldberger and M.~B.~Wise,
  Phys.\ Rev.\ D {\bf 60} (1999) 107505
  [hep-ph/9907218].


\bibitem{Das:2013lqa}
  A.~Das and S.~SenGupta,
  arXiv:1303.2512 [hep-ph].

\bibitem{Choudhury:2014sua}
  S.~Choudhury,
  arXiv:1406.7618 [hep-th].

\bibitem{Davoudiasl:1999jd}
  H.~Davoudiasl, J.~L.~Hewett and T.~G.~Rizzo,
  Phys.\ Rev.\ Lett.\  {\bf 84} (2000) 2080
  [hep-ph/9909255].

\bibitem{ATLAS:2013mma}
  [ATLAS Collaboration],
  ATLAS-CONF-2013-014.

\bibitem{Aad:2012tfa}
  G.~Aad {\it et al.}  [ATLAS Collaboration],
  Phys.\ Lett.\ B {\bf 716} (2012) 1
  [arXiv:1207.7214 [hep-ex]].

\bibitem{Chatrchyan:2013lba}
  S.~Chatrchyan {\it et al.}  [CMS Collaboration],
  JHEP {\bf 1306} (2013) 081
  [arXiv:1303.4571 [hep-ex]].

\bibitem{Brigante:2007nu}
  M.~Brigante, H.~Liu, R.~C.~Myers, S.~Shenker and S.~Yaida,
  Phys.\ Rev.\ D {\bf 77} (2008) 126006
  [arXiv:0712.0805 [hep-th]].

\bibitem{Cremonini:2011iq}
  S.~Cremonini,
  Mod.\ Phys.\ Lett.\ B {\bf 25} (2011) 1867.

\bibitem{Brigante:2008gz}
  M.~Brigante, H.~Liu, R.~C.~Myers, S.~Shenker and S.~Yaida,
  Phys.\ Rev.\ Lett.\  {\bf 100} (2008) 191601.

\bibitem{Buchel:2009tt}
  A.~Buchel and R.~C.~Myers,
  JHEP {\bf 0908} (2009) 016.

\bibitem{Ge:2008ni}
  X.~-H.~Ge, Y.~Matsuo, F.~-W.~Shu, S.~-J.~Sin and T.~Tsukioka,
  JHEP {\bf 0810} (2008) 009.

\bibitem{Ge:2009eh}
  X.~-H.~Ge and S.~-J.~Sin,
  JHEP {\bf 0905} (2009) 051.

\bibitem{Hofman:2009ug}
  D.~M.~Hofman,
  Nucl.\ Phys.\ B {\bf 823} (2009) 174.



\bibitem{Rizzo:2004rq}
  T.~G.~Rizzo,
  JHEP {\bf 0501} (2005) 028
  [hep-ph/0412087].

\bibitem{Dotti:2007az}
  G.~Dotti, J.~Oliva and R.~Troncoso,
  Phys.\ Rev.\ D {\bf 76} (2007) 064038
  [arXiv:0706.1830 [hep-th]].

\bibitem{Torii:2005xu}
  T.~Torii and H.~Maeda,
  Phys.\ Rev.\ D {\bf 71} (2005) 124002
  [hep-th/0504127].

\bibitem{Konoplya:2010vz}
  R.~A.~Konoplya and A.~Zhidenko,
  Phys.\ Rev.\ D {\bf 82} (2010) 084003
  [arXiv:1004.3772 [hep-th]].

\bibitem{Nojiri:2010wj}
  S.~'i.~Nojiri and S.~D.~Odintsov,
  Phys.\ Rept.\  {\bf 505} (2011) 59
  [arXiv:1011.0544 [gr-qc]].

\bibitem{Gherghetta:2010cj}
  T.~Gherghetta,
  arXiv:1008.2570 [hep-ph].


\bibitem{ATLAS:2011ab}
  G.~Aad {\it et al.}  [ATLAS Collaboration],
  Phys.\ Lett.\ B {\bf 710} (2012) 538
  [arXiv:1112.2194 [hep-ex]].



\bibitem{Chakraborty:2012sd}
  S.~Chakraborty and S.~Sengupta,
  arXiv:1208.1433 [gr-qc].




\end{references}
\end{document}